# Generation of second harmonic in off-diagonal magneto-impedance in Co-based amorphous ribbons

N.A. Buznikov[1], S.S. Yoon[2], L. Jin[1], C.O Kim[1] and C.G. Kim [1,*]

[1] Research Center for Advanced Magnetic Materials, Chungnam National University, Daejeon 305-764, Republic of Korea

[2] Department of Physics, Andong National University, Andong 760-749, Republic of Korea

**Abstract**

The off-diagonal magneto-impedance in Co-based amorphous ribbons was measured using a pick-up coil wound around the sample. The ribbons were annealed in air or in vacuum in the presence of a weak magnetic field. The evolution of the first and second harmonics in the pick-up coil voltage as a function of the current amplitude was studied. At low current amplitudes, the first harmonic dominates in the frequency spectrum of the voltage, and at sufficiently high current amplitudes, the amplitude of the second harmonic becomes higher than that of the first harmonic. For air-annealed ribbons, the asymmetric two-peak behaviour of the field dependences of the harmonic amplitudes was observed, which is related to the coupling between the amorphous phase and surface crystalline layers appearing after annealing. For vacuum-annealed samples, the first harmonic has a maximum at zero external field, and the field dependence of the second harmonic exhibits symmetric two-peak behaviour. The experimental results are interpreted in terms of a quasi-static rotational model. It is shown that the appearance of the second harmonic in the pick-up coil voltage is related to the anti-symmetrical distribution of the transverse field induced by the current. The calculated dependences are in a qualitative agreement with the experimental data.

---

[*] Corresponding author. *E-mail address:* cgkim@cnu.ac.kr

# 1. Introduction

The giant magneto-impedance (GMI) effect is studied extensively during last years. The effect is promising due to its possible use for the development of highly sensitive magnetic-field sensors. The GMI implies a strong dependence of the impedance of a conductor on an external magnetic field and has been observed in a wide variety of soft magnetic materials [1]. Usually, the GMI is measured as the field dependence of the voltage drop across a sample. Another method to detect the field-dependent signal consists in the use of the pick-up coil wound around a conductor [2–5]. This method is based on the cross-magnetization process, which appears since the current induces the axial magnetization variation and, hence, the pick-up coil voltage. The effect has been referred to as the off-diagonal magneto-impedance. It has been demonstrated that the off-diagonal magneto-impedance may be preferable for sensor applications since it generates a linear voltage response with enhanced field sensitivity.

On the other hand, the occurrence of higher harmonics in the magneto-impedance response has been detected and analyzed recently in glass-coated microwires [6], Co-based amorphous wires and fibers [7–17], composite wires [18–22] and multi-layered films [23]. The higher harmonics in the voltage response occur at sufficiently high amplitudes of the exciting current and are ascribed to the magnetization reversal of a part of the sample. This mode is often refereed to as the nonlinear magneto-impedance for brevity and in keeping with the traditions of GMI studies [12–16, 19–22]. Under nonlinear conditions, the field sensitivity of higher harmonics in the voltage response turns out to be much greater as compared to the sensitivity of the first harmonic, which is promising for applications [6, 22].

Recently, the generation of the second harmonic in the off-diagonal magneto-impedance in Co-based amorphous ribbons annealed in air in the presence of a weak magnetic field has been reported [24]. In this paper, we present the results of the study of the nonlinear off-diagonal magneto-impedance in Co-based amorphous ribbons annealed in air or in vacuum. The evolution of the first and second harmonics in the pick-up coil voltage with an increase of the current amplitude is analysed. It is demonstrated that formation of the surface crystalline layers in air-annealed samples changes drastically the behaviour of the harmonics. The experimental results are described in the framework of a quasi-static rotational model.

## 2. Experimental details

Amorphous $Co_{66}Fe_4Si_{15}B_{15}$ ribbons prepared by the rapid solidification technique were annealed at a temperature of 380 °C during 8 h. A field of 3 Oe was applied along the longitudinal direction of the sample during the annealing. Two types of the ribbons were investigated. The first type of samples was annealed in air, and the second type was vacuum-annealed. The ribbon length, width and thickness were 3 cm, 2 mm and 20 μm, correspondingly.

In experiments, the amplitude $I_0$ of the ac current passing through the ribbon was changed from 1 to 50 mA, and the current frequency $f$ was 100 and 500 kHz. The longitudinal dc magnetic field $H_e$ created by a Helmholtz coil was changed from −30 to 30 Oe. The positive and negative signs of $H_e$ mean that the direction of the external field is parallel and anti-parallel to that of the annealing field. The pick-up coil was wound around the studied ribbon. The coil had 25 turns and was 5.5 mm in length. The amplitudes of the first and second harmonics, $V_1$ and $V_2$, in the pick-up coil voltage were measured by a lock-in amplifier as a function of the current amplitude, frequency and longitudinal dc magnetic field.

## 3. Experimental results

Figure 1 presents the measured field dependences of the first and second harmonic amplitudes for the ribbons annealed in air. At low current amplitudes, the first harmonic dominates in the pick-up coil voltage frequency spectrum, whereas the second harmonic is negligibly small (see figure 1(*a*)). With a growth of the current amplitude, the contribution of the second harmonic to the voltage response increases (see figure 1(*b*)). At sufficiently high current amplitudes, the amplitude of the second harmonic becomes much higher than that of the first harmonic (see figure 1(*c*)). The dependences of the harmonic amplitudes exhibit asymmetric two-peak behaviour, with the peak at the positive field being higher than the peak at the negative field. It follows from figure 1 that the asymmetry in the field dependence of the second harmonic amplitude is more pronounced in comparison with the first harmonic. The field sensitivity of the second harmonic at $I_0 = 50$ mA is sufficiently high and attains 10 mV Oe$^{-1}$ for the region of positive external fields.

Shown in figure 2 are the field dependences of the harmonic amplitudes measured for the vacuum-annealed ribbons. The first harmonic exhibits single-peak behaviour with the peak at zero external field. The first harmonic amplitude differs from zero within the range of low external field and is less by one order of the magnitude than that in the case of the air-annealed ribbon. The field dependence of the second harmonic amplitude $V_2$ has a symmetric two-peak behaviour, and the contribution of $V_2$ to the pick-up coil voltage increases with the current amplitude.

## 4. Discussion

The observed asymmetric field dependences of the harmonic amplitudes in air-annealed ribbons can be understood as follows. It is well known that the Co-based amorphous ribbons annealed in air in the presence of a weak magnetic field show the asymmetric GMI response measured as the voltage drop across the sample [25–31]. At sufficiently low frequencies, the GMI profile exhibits a drastic step-like change in the impedance near the zero field (so called 'GMI valve'), whereas at high frequencies, the field dependence of the impedance shows asymmetric two-peak behaviour [29]. Since the ribbons field-annealed in vacuum did not display the asymmetry [25, 28], the asymmetric GMI has been attributed to the surface crystallization due to oxidation. It has been confirmed by the x-ray diffraction spectra that the crystalline layer is developed on the surface of the ribbon annealed in air at 380 °C [26, 28]. The composition-depth profile measured by Auger electron spectroscopy has shown that the Co-rich crystalline layer is developed between the oxidation layer and the amorphous bulk due to the diffusion of B to the surface to form oxidation [26]. Furthermore, the differential thermal analysis profile of the ribbon annealed in air has shown that the surface layer has the crystallization temperature about 330 °C, much lower than the crystallization temperature of the amorphous bulk of about 540 °C [27]. The existence of the surface crystalline phase has also been confirmed in the study of the surface magnetic properties of the ribbons by means of the magneto-optical Kerr effect [30]. The analysis of the experimental data for the samples with surface layers removed by etching has shown that the thickness of the surface crystalline layer is of the order of 1 μm [30]. It has been suggested that the formation of a hard magnetic phase at the ribbon surface results in the asymmetry in the field dependence of the impedance due to coupling between the amorphous bulk and surface crystalline layer [29, 31].

In order to interpret the off-diagonal magneto-impedance in air-annealed ribbons, a rotational model is proposed. It is assumed that the ribbon consists of the amorphous bulk and two surface crystalline layers [32]. The crystalline layers have the unidirectional anisotropy induced by the annealing field. The value of the unidirectional anisotropy field $H_u$ in the crystalline layers is sufficiently high and attains several hundreds of Oersted [27, 29]. The unidirectional anisotropy field in the outer layers makes the angle $\varphi$ with respect to the transverse direction (see figure 3). The deviation of the unidirectional anisotropy field $H_u$ from the annealing field direction may be attributed to the influence of the amorphous bulk on the surface crystallization process [32–34]. It is assumed also that these surface layers have different thickness, $d_1$ and $d_2$. The difference in the thicknesses of the crystalline layers may be related to the peculiarities of the annealing process. During the annealing, one side of the ribbon remained in the air and the other was located on the substrate, which may result in the difference in the thicknesses of the crystalline layers.

It is assumed that the amorphous bulk has a single-domain structure, the uniaxial anisotropy field $H_a$ is constant, and the anisotropy axis makes the angle $\psi$ with the transverse direction (see figure 3). The coupling between amorphous and crystalline phases induces the effective bias field $H_b$ in the amorphous bulk. The bias field is anti-parallel to the unidirectional anisotropy field $H_u$ [32] and is assumed, for simplicity, to be constant.

Since in the experiments the current frequency is comparatively low, the skin effect in the ribbon can be neglected. Therefore, the distribution of the transverse ac magnetic field $H_1$ induced by the current can be expressed as

$$H_1(x,t) = (2x/D)H_0 \sin(2\pi f t), \qquad (1)$$

where $H_0 = 2\pi I_0/cw$ is the amplitude of the transverse field, $x$ is the coordinate perpendicular to the ribbon plane, $x=0$ corresponds to the central plane of the ribbon, $w$ and $D$ are the ribbon width and thickness, respectively, and $c$ is the velocity of light.

The variation of the magnetization induced by the current results in the appearance of a voltage $V$ in the pick-up, which can be expressed through the longitudinal magnetization component $M_z$:

$$V = -\frac{4\pi w N}{c} \times \int_{-D/2}^{D/2} \frac{\partial M_z}{\partial t} dx, \qquad (2)$$

where $N$ is the number of turns in the pick-up coil.

Since the unidirectional anisotropy field $H_u$ in the crystalline layers is high, the appearance of the pick-up coil voltage response is related to changes in the magnetization in the amorphous bulk only. The variation of the magnetization under the effect of the ac current can be described in the framework of the quasi-static approximation and can be found by minimizing the free energy. Under the assumptions made above, the free energy density $U$ can be presented as a sum of the anisotropy, bias and Zeeman terms:

$$U = (MH_a/2)\sin^2(\theta - \psi) + MH_b\cos(\theta - \varphi) - MH_e\sin\theta - MH_I\cos\theta. \qquad (3)$$

Here $M$ is the saturation magnetization and $\theta$ is the equilibrium angle between the magnetization vector and the transverse direction.

It should be noted that similar rotational models have been developed previously for the wire geometry to describe the second harmonic signal. The second harmonic in the pick-up coil voltage has been calculated for the case of the circular anisotropy in the wire, $\psi=0$, and the absence of the bias field, $H_b=0$ [6, 13, 21]. The influence of the deviation of the anisotropy axis from the circular direction on the second harmonic in the voltage measured across the sample ends has been analyzed neglecting the spatial distribution of the field induced by the ac current [7–9] and taking into account the effect of torsional stress [10, 11, 15].

The distribution of the longitudinal, $M_z(x,t)=M\sin\theta$, and transverse, $M_y(x,t)=M\cos\theta$, magnetization components are given by the minimum conditions of the free energy $\partial U/\partial \theta=0$ and $\partial^2 U/\partial\theta^2>0$, which results in

$$H_a M_y M_z \cos(2\psi) - H_a(M_y^2 - M_z^2)\sin(2\psi)/2$$
$$- MM_y(H_e - H_b \sin\varphi) + MM_z(H_1 - H_b \cos\varphi) = 0,$$
$$H_a \cos(2\psi)(M_y^2 - M_z^2) + 2H_a M_y M_z \sin(2\psi)$$
$$+ MM_y(H_1 - H_b \cos\varphi) + MM_z(H_e - H_b \sin\varphi) > 0. \tag{4}$$

Using equations (2) and (4), we find for the pick-up coil voltage

$$V = V_0(H_0/M)\cos(2\pi ft) \times \left[\int_0^{\xi_1} \frac{M_y M_z \xi d\xi}{P_1} + \int_{\xi_2}^0 \frac{M_y M_z \xi d\xi}{P_2}\right], \tag{5}$$

where

$$P_1 = (H_a/M)[\cos(2\psi)(M_y^2 - M_z^2) + 2M_y M_z \sin(2\psi)]$$
$$+ M_z(H_e - H_b \sin\varphi) + M_y[H_0 \xi \sin(2\pi ft) - H_b \cos\varphi],$$
$$P_2 = (H_a/M)[\cos(2\psi)(M_y^2 - M_z^2) + 2M_y M_z \sin(2\psi)] \tag{6}$$
$$+ M_z(H_e - H_b \sin\varphi) - M_y[H_0 \xi \sin(2\pi ft) + H_b \cos\varphi],$$

$V_0 = 4\pi^2 NMfwD/c$, $\xi = 2x/D$, $\xi_1 = 1 - 2d_1/D$, $\xi_2 = -1 + 2d_2/D$ and the magnetization components $M_z$ and $M_y$ are given by equations (4).

The calculated pick-up coil voltage $V$ is shown in figure 4 as a function of time $t$ for the fixed external field $H_e$ and different current amplitudes $I_0$. At $I_0 = 1$ mA the dependence of $V$ on $t$ has the period of the variation of the ac current. With the increase of $I_0$, the behaviour of the pick-up voltage changes drastically. It follows from figure 4 that at sufficiently high $I_0$, the dependence $V(t)$ transforms to the function with a period half of that of the ac current, which denotes that the second harmonic becomes dominant in the frequency spectrum.

The evolution of the pick-up coil voltage with the increase of the current amplitude can be explained as follows. At low current amplitude, when the last term in the right-hand part of equation (3) is very small, the magnetization variations in two parts of the ribbon, $x > 0$ and $x < 0$, give the opposite contribution to the pick-up coil voltage due to the anti-symmetrical transverse magnetic field distribution over the ribbon thickness. This case corresponds to the linear off-diagonal impedance, when the first harmonic dominates in the pick-up coil voltage spectrum, and the voltage response can be described in terms of the field and frequency dependent surface impedance tensor [2, 4, 5]. It should be noted that the first harmonic in the pick-up coil voltage appears only if the surface crystalline layers have different thickness [35]. The voltage response is proportional to the difference in the surface layer thickness, $|d_1-d_2|$, since the contributions from the other parts of the ribbons to the pick-up coil voltage are compensated completely. With a growth of the current amplitude, the anti-symmetrical transverse magnetic field distribution results in the difference in the magnetization variations in two parts of the ribbon, $x>0$ and $x<0$. The difference increases with the current amplitude, which leads to the growth of the second harmonic contribution to the pick-up coil voltage. Note that in contrast to the first harmonic, the second harmonic amplitude is non-zero even in the case of the same thickness of the surface layers, $d_1=d_2$.

The amplitudes of the first and second harmonics can be found by means of the Fourier transformation of equation (5). The comparison of the field dependences of the calculated harmonic amplitudes, $V_1$ and $V_2$, with the experimental results is shown in figure 1. The second harmonic increases with the current amplitude $I_0$ approximately as $I_0^2$, and the second harmonic becomes dominant in the pick-up coil voltage at high current amplitudes. The calculated field dependences of the first and second harmonic amplitudes describe qualitatively the main features of the experimental data. The disagreements between the calculations and experimental data are attributed to approximations made in the model. In particular, we neglect a domain structure and the coordinate dependences of the bias field and the anisotropy field. Taking into account the spatial distribution of the bias field may be essential for a detailed explanation of the evolution of the pick-up coil voltage frequency spectrum with the change in the current amplitude. Moreover, further experimental investigation of the effect of the annealing conditions on the thickness of the surface crystalline layers and on the off-diagonal magneto-impedance should be carried out.

In the vacuum-annealed ribbons, the surface crystalline layers do not appear [25, 28]. Therefore, to calculate the pick-up coil voltage in this case we should assume in the expressions above that $d_1=d_2=0$ and $H_b=0$. In the framework of the rotational model, the first harmonic in the pick-up coil voltage equals zero. The calculated field dependence of the second harmonic amplitude $V_2$ for the vacuum-annealed ribbons is shown in figure 2. Since there is no the bias field in the vacuum-annealed samples, the field dependence of $V_2$ has the symmetric two-peak behaviour. It follows from figure 2 that the calculated dependences are in a sufficiently good agreement with the measured field dependences of the second harmonic amplitude.

The appearance of the first harmonic in the pick-up coil voltage in the vacuum-annealed ribbons may be attributed to the effect of the domain-wall motion. Indeed, it follows from figure 2 that the first harmonic amplitude differs from zero only within the range of the low external field, where the domain structure may exist. It is well known that the domain-walls motion do not contribute to the off-diagonal impedance in the case of the perfect transverse domain structure, since the domain-walls motion contribution to the permeability has only diagonal components [36, 37]. If the anisotropy axis deviates from the transverse direction, the contribution from the domain-walls motion to the off-diagonal impedance differs from zero and results in the single-peak field dependence of the impedance. The estimations made in [35] have shown that the domain-walls motion contribution to the off-diagonal magneto-impedance is much less than that to the GMI measured as the voltage drop across the ribbon. In this connection, the first harmonic in the vacuum-annealed ribbons is much less than that in the air-annealed samples, and the first harmonic in the air-annealed ribbons has two-peak field dependence even at sufficiently low frequencies (see figure 1).

In the conclusion of this section, it should be noted that most of the studies of the nonlinear magneto-impedance deal with the generation of higher harmonics in the voltage measured at the sample ends. The occurrence of the second harmonic in the pick-up coil voltage has been studied in detail previously for glass-coated Co-based amorphous microwires [6], Co-based amorphous wires extracted from the melt [13], composite wires consisting of highly conductive core and soft magnetic shell [21] and multi-layered films [23]. For these samples, the second harmonic becomes dominant in the frequency spectrum of the voltage, when the current amplitude exceeds some threshold value, and the nonlinear response is related to the magnetization reversal of part of the sample. The threshold current amplitude, at which the magnetization reversal takes place, can be estimated from the condition of equality of the amplitude of the ac magnetic field induced by the current and the anisotropy field [6]. In the studied amorphous ribbons, the mechanism resulting in the appearance of the second harmonic has a fundamentally different nature. Simple estimations demonstrate that the amplitude $H_0$ of the transverse magnetic field is too low for the magnetization reversal of the ribbon even at $I_0 = 50$ mA, when $H_0 \cong 0.15$ Oe, and the appearance of the second harmonic is attributed to the different magnetization variations in two parts of the ribbon under influence of the transverse magnetic field.

## 5. Conclusions

The off-diagonal magneto-impedance in field-annealed Co-based amorphous ribbons was studied. The effect of the current amplitude on the first and second harmonics in the pick-up coil voltage was investigated. It was observed that the contribution of the second harmonic to the voltage response increases with the current amplitude, and at sufficiently high current amplitudes, the second harmonic becomes dominant. The asymmetric field dependences of the harmonic amplitudes in the ribbons annealed in air are related to the coupling between the amorphous phase and surface crystalline layers appearing after annealing. The generation of the second harmonic in the voltage is attributed to the anti-symmetrical distribution of the field induced by the current. A sufficiently high sensitivity of the second harmonic to the external magnetic field makes the studied nonlinear off-diagonal magneto-impedance promising for the development of sensitive magnetic-field sensors.

**Acknowledgments**

This work was supported by the Korea Science and Engineering Foundation through ReCAMM and by the Korea Electrical Engineering and Science Research Institute. N.A. Buznikov would like to acknowledge the support of the Brain Pool Program.

**References**

[1] Knobel M, Vazquez M and Kraus L 2003 *Handbook of Magnetic Materials* vol 15, ed K H J Buschow (Amsterdam: Elsevier) p 497

[2] Usov N A, Antonov A S and Lagar'kov A N 1998 *J. Magn. Magn. Mater.* **185** 159

[3] Antonov A S, Iakubov I T and Lagarkov A N 1998 *J. Magn. Magn. Mater.* **187** 252

[4] Panina L V, Mohri K and Makhnovskiy D P 1999 *J. Appl. Phys.* **85** 5444

[5] Makhnovskiy D P, Panina L V and Mapps D J 2001 *Phys. Rev.* B **63** 144424

[6] Antonov A S, Buznikov N A, Iakubov I T, Lagarkov A N and Rakhmanov A L 2001 *J. Phys. D: Appl. Phys.* **34** 752

[7] Gomez-Polo C, Vazquez M and Knobel M 2001 *Appl. Phys. Lett.* **78** 246

[8] Gomez-Polo C, Vazquez M and Knobel M 2001 *J. Magn. Magn. Mater.* **226–230** 712

[9] Gomez-Polo C, Knobel M, Pirota K R and Vazquez M 2001 *Physica* B **299** 322

[10] Gomez-Polo C, Pirota K R and Knobel M 2002 *J. Magn. Magn. Mater.* **242–245** 294

[11] Losin C, Gomez-Polo C, Knobel M and Grishin A 2002 *IEEE Trans. Magn.* **38** 3087

[12] Duque J G S, de Araujo A E P, Knobel M, Yelon A and Ciureanu P 2003 *Appl. Phys. Lett.* **83** 99

[13] Antonov A S, Buznikov N A, Granovsky A B, Perov N S, Prokoshin A F, Rakhmanov A A and Rakhmanov A L 2003 *Sensors Actuators* A **106** 213

[14] Clime L, Rudkowska G, Duque J G S, de Araujo A E P, Knobel M, Ciureanu P and Yelon A 2004 *Physica* B **343** 410

[15] Duque J G S, Gomez-Polo C, Yelon A, Ciureanu P, de Araujo A E P and Knobel M 2004 *J. Magn. Magn. Mater.* **271** 390

[16] Gomez-Polo C, Duque J G S and Knobel M 2004 *J. Phys.: Cond. Mater.* **16** 5083

[17] Chen A P, Britel M R, Zhukova V, Zhukov A, Dominguez L, Chizhik A B, Blanco J M and Gonzalez J 2004 *IEEE Trans. Magn.* **40** 3368

[18] Beach R S, Smith N, Platt C L, Jeffers F and Berkowitz A E 1996 *Appl. Phys. Lett.* **68** 2753

[19] Kurlyandskaya G V, Yakabchuk H, Kisker E, Bebenin N G, Garcia-Miquel H, Vazquez M and Vas'kovskiy V O 2001 *J. Appl. Phys.* **90** 6280

[20] Kurlyandskaya G V, Kisker E, Yakabchuk H and Bebenin N G 2002 *J. Magn. Magn. Mater.* **240** 206

[21] Antonov A S, Buznikov N A, Granovsky A B, Iakubov I T, Prokoshin A F, Rakhmanov A L and Yakunin A M 2002 *J. Magn. Magn. Mater.* **249** 315

[22] Kurlyandskaya G V, Garcia-Arribas A and Barandiaran J M 2003 *Sensors Actuators* A **106** 239

[23] Buznikov N A, Antonov A S, D'yachkov A L and Rakhmanov A A 2004 *J. Phys. D: Appl. Phys.* **37** 518

[24] Buznikov N A, Yoon S S, Kim C O and Kim C G 2005 *IEEE Trans. Magn.* **41** 3646

[25] Kim C G, Jang K J, Kim H C and Yoon S S 1999 *J. Appl. Phys.* **85** 5447

[26] Jang K J, Kim C G, Yoon S S and Shin K H 1999 *IEEE Trans. Magn.* **35** 3889

[27] Rheem Y W, Kim C G, Kim C O and Yoon S S 2002 *J. Appl. Phys.* **91** 7433

[28] Kim C G, Kim C O, Yoon S S, Stobiecki T and Powroznik W 2002 *J. Magn. Magn. Mater.* **242–245** 467

[29] Kim C G, Kim C O and Yoon S S 2002 *J. Magn. Magn. Mater.* **249** 293

[30] Kim C G, Rheem Y W, Kim C O, Shalyguina E E and Ganshina E A 2003 *J. Magn. Magn. Mater.* **262** 412

[31] Rheem Y W, Jin L, Yoon S S, Kim C G and Kim C O 2003 *IEEE Trans. Magn.* **39** 3100

[32] Buznikov N A, Kim C G, Kim C O and Yoon S S 2005 *J. Magn. Magn. Mater.* **288** 205

[33] Buznikov N A, Kim C G, Kim C O and Yoon S S 2004 *Appl. Phys. Lett.* **85** 3507

[34] Yoon S S, Buznikov N A, Kim D Y, Kim C O and Kim C G 2005 *Eur. Phys. J.* B **45** 231

[35] Buznikov N A, Kim C G, Kim C O, Jin L and Yoon S S 2005 *J. Appl. Phys.* **98** 113908

[36] Panina L V, Mohri K, Uchiyama T, Noda M and Bushida K 1995 *IEEE Trans. Magn.* **31** 1249

[37] Makhnovskiy D P and Panina L V 2000 *Sensors Actuators* A **81** 91

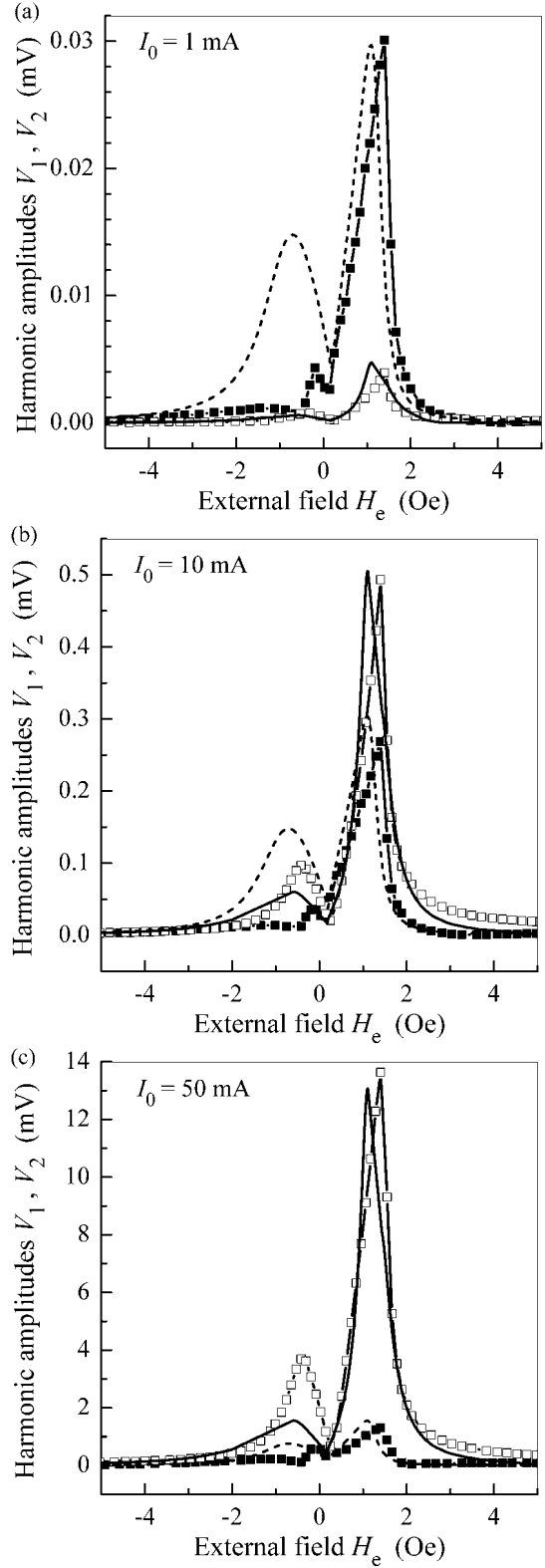

**Figure 1.** The dependences of the harmonic amplitudes on the external field for air-annealed ribbon at $f=100$ kHz and different current amplitudes $I_0$: (*a*) $I_0=1$ mA; (*b*) $I_0=10$ mA; (*c*) $I_0=50$ mA. Symbols – experimental data (■, first harmonic; □, second harmonic); lines – calculations (dashed lines, first harmonic; solid lines, second harmonic). Parameters used for calculations are $M=600$ G, $H_a=1$ Oe, $H_b=0.25$ Oe, $\psi=0.05\pi$, $\varphi=0.35\pi$, $d_1=1$ μm, $d_2=0.6$ μm.

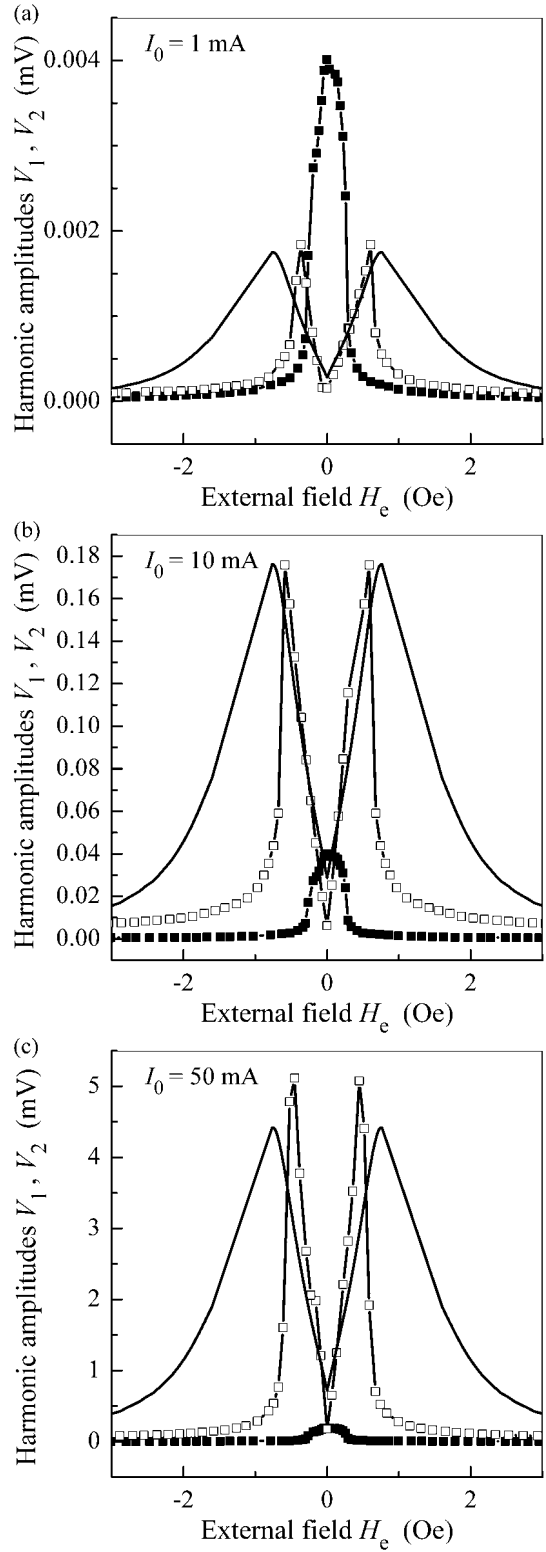

**Figure 2.** The measured dependences of the harmonic amplitudes on the external field for vacuum-annealed ribbon at $f=100$ kHz and different current amplitudes $I_0$: (*a*) $I_0=1$ mA; (*b*) $I_0=10$ mA; (*c*) $I_0=50$ mA (■, first harmonic; □, second harmonic). Solid lines – the calculated dependences of the second harmonic amplitude. Parameters used for calculations are $M=600$ G, $H_a=0.9$ Oe, $\psi=0.05\pi$.

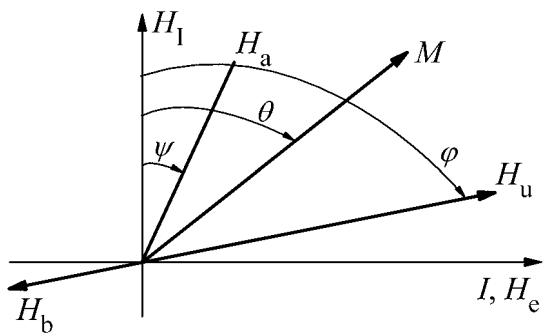

**Figure 3.** A sketch of the coordinate system used for analysis. All the vectors lie within the ribbon plane.

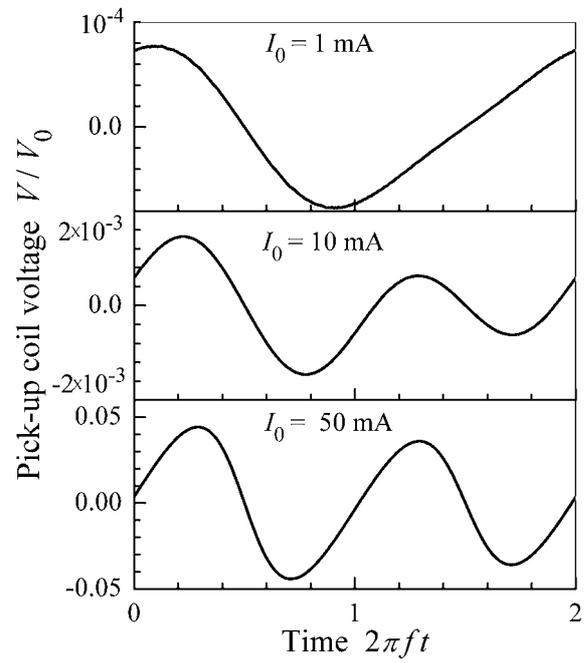

**Figure 4.** The calculated pick-up coil voltage for air-annealed ribbon versus dimensionless time at $H_e=0.25$ Oe. Parameters used for calculations are $M=600$ G, $H_a=1$ Oe, $H_b=0.25$ Oe, $\psi=0.05\pi$, $\varphi=0.35\pi$, $d_1=1$ μm, $d_2=0.6$ μm.